\documentclass[11pt]{article}
\providecommand{\keywords}[1]{\textbf{\textit{Keywords$\colon$}} #1}
\usepackage[top=1in, bottom=1in, left=1in, right=1in]{geometry}
\usepackage{setspace}
\usepackage{multirow}
\usepackage[section]{placeins}
\usepackage{subfig}
\usepackage{geometry}
\usepackage{placeins}
\usepackage[export]{adjustbox}
\usepackage{blkarray}
\usepackage{placeins}
\usepackage[hidelinks]{hyperref}
\usepackage{hyperref}
\usepackage{float}
\usepackage{rotating}
\usepackage{amsmath, amsthm, amssymb, lipsum}
\usepackage{comment}
\usepackage{calc}
\usepackage[T1]{fontenc}
\usepackage{multirow}
\usepackage{framed}
\usepackage{graphicx}
\usepackage{pdflscape}
\usepackage{hhline}
\usepackage{array}
\newcolumntype{L}[1]{>{\raggedright\let\newline\\\arraybackslash\hspace{0pt}}m{#1}}
\newcolumntype{C}[1]{>{\centering\let\newline\\\arraybackslash\hspace{0pt}}m{#1}}
\newcolumntype{R}[1]{>{\raggedleft\let\newline\\\arraybackslash\hspace{0pt}}m{#1}}
\usepackage{color}
\usepackage{array}
\usepackage{amsthm}
\usepackage[fleqn]{mathtools}
\usepackage{caption} 
\usepackage{amsmath} 
\newcommand\numberthis{\addtocounter{equation}{1}\tag{\theequation}}
\usepackage{enumerate}
\usepackage[utf8]{inputenc}
\usepackage{fancyhdr}
\usepackage{breqn}
\usepackage{graphics} 

\usepackage{amsmath} 
\usepackage{amssymb} 
\usepackage{epstopdf}
\usepackage{lscape}
\usepackage[title]{appendix}
\usepackage{epsfig} 
\renewcommand{\qedsymbol}

\title{\LARGE \bf
Markovian Analysis of Coordination Strategies in Tandem Polling Queues with Setups
}
\author{
Ravi Suman$^1$*, Ananth Krishnamurthy$^2$\\Industrial and Systems Engineering Department\\University of Wisconsin-Madison, Madison, WI-53726, USA}
\begin{document}
\begin{center}
\vspace{1in} \Large{\textbf{Markovian Analysis of Coordination Strategies in Tandem Polling Queues with Setups}}\\
\vspace{0.1in}
\large{Ravi Suman$^{1*}$, Ananth Krishnamurthy$^2$}\\
$^1$University of Wisconsin-Madison\\%
$^2$Indian Institute of Management Bangalore\\%
Email: rsuman@wisc.edu, ananthk@iimb.ac.in\\
\end{center}
\begin{abstract}
We analyze a network of tandem polling queues with two stations operating under synchronized polling (SP) and out-of-sync polling (OP) strategies, and with nonzero setups. We conduct an exact analysis using a decomposition approach to compare the performance in terms of throughput and mean waiting times to investigate when one strategy might be preferred over the other. We also numerically investigate the condition for network stability operating under the two strategies and show that polling network is unstable when there is bottleneck at downstream stations. We find that the SP strategy outperforms the OP strategy in case of product and station symmetric networks while under certain settings of product and station asymmetry, OP strategy outperforms the SP strategy.\\

\noindent\keywords{Polling Queues, Multi-products, Multi-station, Decomposition, Polling Strategies}
\end{abstract}

\section{Introduction}\label{Introduction}
Polling queues find applications when multiple products compete for a common resource. In a polling queue, a single server serves multiple queues of products, visiting the queues one at a time in a fixed cyclic manner. In manufacturing, polling queues have been used to model flow of multiple products undergoing manufacturing operations in a factory. In healthcare, polling queues have been used to model the flow of different types of patients through various activities in a hospital or clinic. In transportation, polling queues have been used to model multiple traffic flows in a transportation network. Comprehensive survey on the analysis of polling queues can be found in $\left(\text{Takagi }\cite{Takagi2000}, \text{Vishnevskii \& Semenova }\cite{Vishnevskii2006}\right)$.\\

While a majority of existing research on polling queues focus on the single-station polling queue, this work focuses on the analysis of coordination strategies in tandem network of polling queues with setups. In tandem networks of polling queues, coordination of service across different stations is essential for achieving efficient network performance in terms of throughput, queue lengths, and waiting times. One extreme, among the spectrum of coordination mechanism, is the \emph{Independent polling} $(IP)$ strategy, where each polling station acts independently and serves customers of various types based on process characteristics, arrival patterns at that particular station alone. Another strategy would be to synchronize the operations at consecutive polling queues so that the same type of customer is being processed at the two stations. For example, rolled aluminum plates need to be stretched soon after they have been heat treated, before they cool down. This necessitates that both stations process a particular type of alloy at the same time in their respective cycles. In this case, these operations can be modeled as tandem polling queues operating under \emph{synchronous polling} $(SP)$  strategy. Another possibility is that downstream operations are deliberately kept out of sync with the upstream operations to improve the flow. For example, it may be desired that while the saws are processing one type of alloy, the downstream inspection station is setup to inspect the parts after they got complete at the saw. In this case, these operations can be modeled as tandem polling queues operating under an \emph{out-of-sync} $(OS)$ strategy. Between these extremes, and in addition to these specific strategies, several other coordination strategies are possible. Further, such coordination can be both essential and preferred in application of tandem polling queues to healthcare and transportation settings.\\

Despite the importance of tandem network of polling queues, there has been limited studies of such networks. Exact analysis of polling models is only possible in some cases, and even then numerical techniques are usually required to obtain waiting times at each queue.  Our research focus aims to investigate when such coordination will be beneficial in a network. We believe that when there are significant differences between product types (in terms of service times and demands) and stations (in terms of switchover times and traffic intensities); designing and operating a tandem polling network with suitable coordination strategy can significantly improve its performance. We propose a decomposition approach and analyze the decomposed system using matrix-geometrics approach to analyze the steady state of the system. We show that this approach reduces computational complexity and provides reasonable accuracy in performance estimation. Second, we investigate the impact of different manufacturing settings, such as, location of bottleneck stations, asymmetry in waiting times, and setup times on systems performance measures. We find that the location of bottleneck station and differences in service rates can have significant impact on the waiting times.\\

The rest of the paper is organized as follows. In Section \ref{LiteratureReview7}, we provide a brief literature review on polling queues and analysis of tandem network of queues. We describe the system in Section \ref{SystemDescription7} and the approach used to analyze the two-station system in Section \ref{QueueingAnalysis7}. In Section \ref{NumericalResults7}, we validate our approach and provide useful numerical insights. Finally, we conclude and provide future extensions in Section \ref{Conclusions7}.
\section{Literature Review}\label{LiteratureReview7}
Polling queues and their applications have been an active field of research for the past few decades. Takagi \cite{Takagi2000}, Vishnevskii  and Semenova \cite{Vishnevskii2006}, and  Boona et al. \cite{Boona11} provide a comprehensive survey on polling queues and their applications. We group our discussion of the literature in three categories$\colon$ polling queue with zero setups, polling queue with non-zero setups, and network of polling queues.\\

\textbf{Polling queue with zero setups}$\colon$ One of the earliest techniques for analyzing polling queues with zero setups uses a \emph{server vacation model}, where the server periodically leaves a queue and takes a vacation to serve other queues. Fuhrmann et al. \cite{Fuhrmann85} uses such a vacation model to study a symmetric polling station with $Q$ queues served in a cyclic order by a single server and determines the expressions for sojourn times under exhaustive, gated, and $k$-limited service discipline. They show that the stationary number of customers in a single station polling queue (summed over all the queues) can be written as the sum of three independent random variables$\colon\left(i\right)$ the stationary number of customers in a standard M/G/I queue with a dedicated server, $\left(ii\right)$ the number of customers in the system when the server begins an arbitrary vacation (changeover), and $\left(iii\right)$ number of arrivals in the system during the changeover. Boxma et al. \cite{Boxma87} use a stochastic decomposition to estimate the amount of work (time needed to serve a specific number of customers) in cyclic-service systems with hybrid service strategies (e.g., semi-exhaustive for first product class, exhaustive for second and third product class, and gated for remaining product classes) and use the decomposition results to obtain a pseudo-conservation law for such cyclic systems.\\

\textbf{Polling queue with non-zero setups}$\colon$ Several studies have used transform methods to find the distributions for waiting times, cycle times, and queue lengths in a single-station polling queue with setups. Cooper et al. \cite{RBCooper96} propose a decomposition theorem for polling queues with non-zero switchover times and show that the mean waiting times is the sum of two terms$\colon\left(\text{1}\right)$ the mean waiting time in a "corresponding" model in which the switchover times are zero, and $\left(\text{2}\right)$ a simple term that is a function of mean switchover times. Srinivasan et al. \cite{Srinivasan95} use Laplace–Stieltjes Transform $\left(\text{LST}\right)$ methods to compute the moments of the waiting times in $R$ polling queues with nonzero-setup-times for exhaustive and gated service. The algorithm proposed requires estimation of parameters with $\log{\left(R\mathcal{E}\right)}$ complexity, with $\mathcal{E}$ as the desired level of accuracy. Once the parameters have been calculated, mean waiting times may be computed with $\mathcal{O}\left(R\right)$ elementary operations. Borst and Boxma \cite{Borst97} generalize the approach used by Srinivasan et al. \cite{Srinivasan95} to derive the joint queue length distribution for any service policy. Boxma et al. \cite{Boxma09} analyzes a polling system of $R$-queues with setup times operating under gated policy and determine the LST for cycle times under different scheduling disciplines such as FIFO and LIFO. They show that LST of cycle times is only dependent on the polling discipline at each queue and is independent of the scheduling discipline used within each queue.\\

In addition to LST techniques, mean value analysis has also been used to estimate performance measures for polling queues with nonzero setups. Hirayama et al. \cite{Hirayama04} developed a method for obtaining the mean waiting times conditioned on the state of the system at an arrival epoch. Using this analysis, they obtain a set of linear functional equations for the conditional waiting times. By applying a limiting procedure, they derive a set of $R(R +1)$ linear equations for the unconditional mean waiting times, which can be solved in $\mathcal{O}\left(R^6\right)$ operations. Winands et al. \cite{Winands06} calculates the mean waiting times in a single-station multi-class polling queue with setups for both exhaustive and gated service disciplines. They use mean value analysis to determine the mean waiting times at the polling queue. They derive a set of $R^{2}$ and $R\left(R +1\right)$ linear equations for waiting time figures in case of exhaustive and gated service. In these studies of polling queues using LST techniques or mean value analysis, the authors have restricted their scope of study to single-station polling queues. Extending their approach to tandem network of polling queue will increase the computational complexity. Therefore, in our work, we propose a decomposition based approach.\\

$\textbf{Network of polling queues}\colon$ Altman and Yechiali \cite{Altman94} study a closed queueing network for token ring protocols with $Q$ polling stations, where a product upon completion of the service is routed to another queue probabilistically. They determine explicit expressions for the probability generating function for the number of products at various queues. However, the system considered is closed system with $N$ products in circulation, which could be a restrictive assumption in some applications. Jennings \cite{Jennings08} conducts a heavy traffic analysis of two polling queues for two stations in series and prove limit theorems for exhaustive and gated discipline for the diffusion scaled, two-dimensional total workload process using heavy traffic analysis. Suman and Krishnamurthy (\cite{Suman18} -- \cite{Suman22}) study a two-product two-station tandem network of polling queues with finite buffers using Matrix-Geometric approach. However, the analysis is restricted to systems with small buffer capacity. In comparison, this paper analyzes an open network of two polling queues with exogenous arrivals using decomposition.
\section{System Description and Overview of Approach}\label{SystemDescription7}
In this section, we describe the system and provide an overview of the approach to estimate performance measures for the system.
\subsection{System Description}\label{System Description}
We analyze a two-product two-station tandem polling queue operating under two different polling strategies$\colon$ 1) \emph{Synchronous polling} $(SP)$ and 2) \emph{Out-of-sync polling} $(OP)$. There is an exogenous arrival of product of type $i$ for $i ={}1, 2$ to their respective queue at station 1 according to independent Poisson process with parameter $\lambda_i$. In the two polling strategies, each product type is served by a single server at station $j$ for $j ={}1, 2$ in a fixed cyclic order. After service at station 1, products move from station 1 to station 2, and exit the system after the service is completed at station 2. Service times at these stations for product type $i$ are iid exponential random variables with parameter $\mu_{i1}$ and $\mu_{i2}$ at station 1 and 2 respectively. We assume a common setup time for product $i$ at station 1 and station 2 when a server switches from queue $i'$ to queue $i$, for $i' ={}1, 2$ and $i' \neq i$. We denote this setup time as $H_{i}$ that has an exponential distribution with rate $\mu_{s_{i}}$. We assume that the setups are state independent for the queues at station 1, i.e., the server incurs a setup time at the polled queue whether or not products are waiting at the queue. We also assume that setup times are independent of service times and queue type. Let $N_{1}$ and $N_{2}$ be the respective maximum number of type 1 and type 2 products allowed to queue at both  stations. We assume that an arriving type $i$ product is lost if the queue length of its type at that station is $N_{i}$ (including the product being served) at the time of arrival. Next, we summarize the differences between the two polling strategies below.\\

\begin{figure}[H]%
    \centering
    \subfloat[Both servers are serving type 1 products.]{{\includegraphics[width=7.5cm, frame]{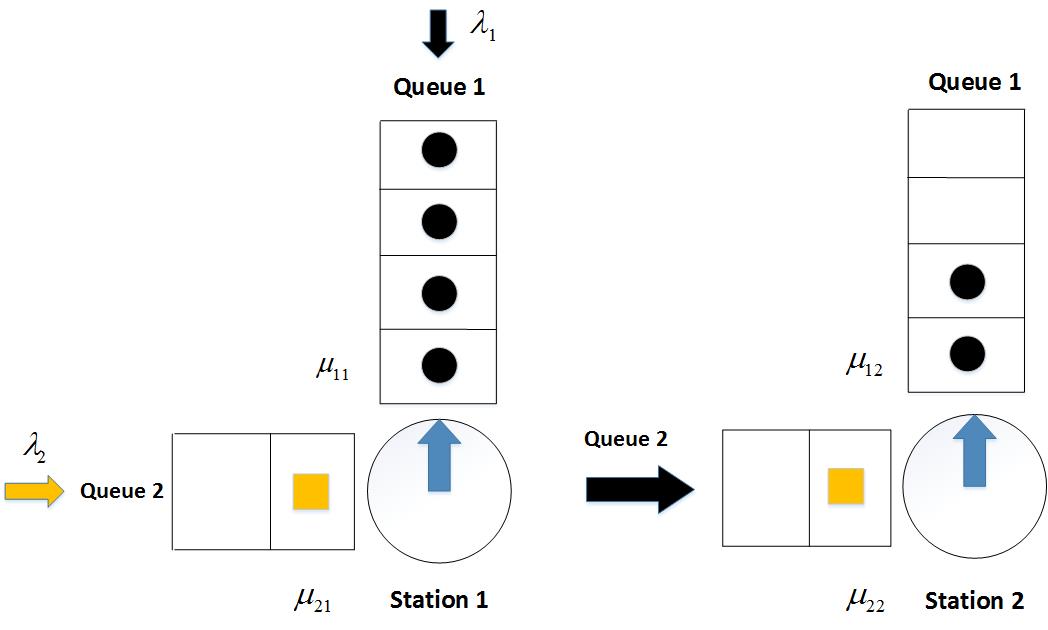}}\label{fig:mesh7.1.1}}%
    \qquad
    \subfloat[Both servers are serving type 2 products.]{{\includegraphics[width=7.5cm, frame]{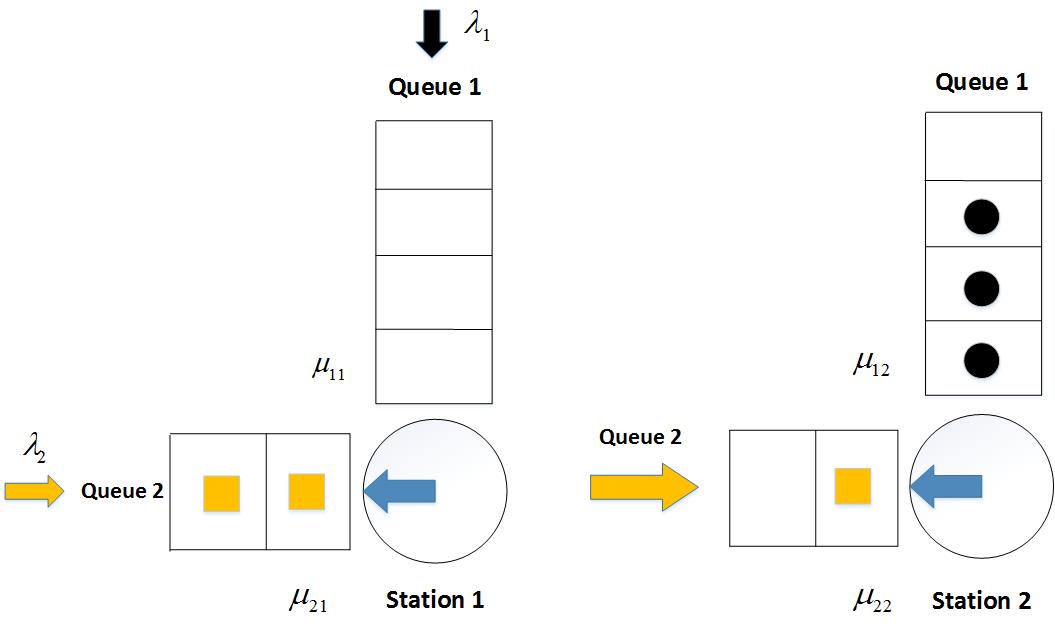}}\label{fig:mesh7.1.2}}%
    \caption{Illustration of synchronous polling strategy (SP).}%
    \label{fig:mesh7.1}%
\end{figure}

\noindent \textbf{Synchronous polling strategy (SP)}$\colon$ In this strategy, station 1 operates under exhaustive policy, but the server at station 2 always serves the same product type being served at station 1. In other words, if the server at station 1 switches to perform setup for products of type 2 after emptying the queue of product type 1, the server at station 2 also switches to perform setup for products of type 2 even if its queue for type 1 product is not empty. However, if station 1 is empty (in both the queues), station 2 continues to serve its queue until its queue is  empty, or until station 1 begins to perform setup for a queue again after the arrival of a product. In this event, the server at station 2 switches to setup for the same product type being setup for at station 1. Figure $\ref{fig:mesh7.1}$ illustrates a possible scenario under the synchronous polling strategy. Initially (in Figure \ref{fig:mesh7.1.1}), the server at both the stations are serving products of type 1. Subsequently, when the server at station 1 empties its queues, it switches to do setup for products of type 2 and the server at station 2 does the same (in Figure \ref{fig:mesh7.1.2}).\\

\graphicspath {{Figures/}}

\begin{figure}[H]%
    \centering
    \subfloat[Server at station 1 is serving type 1 products while server at station 2 is serving type 2 products.]{{\includegraphics[width=7.5cm, frame]{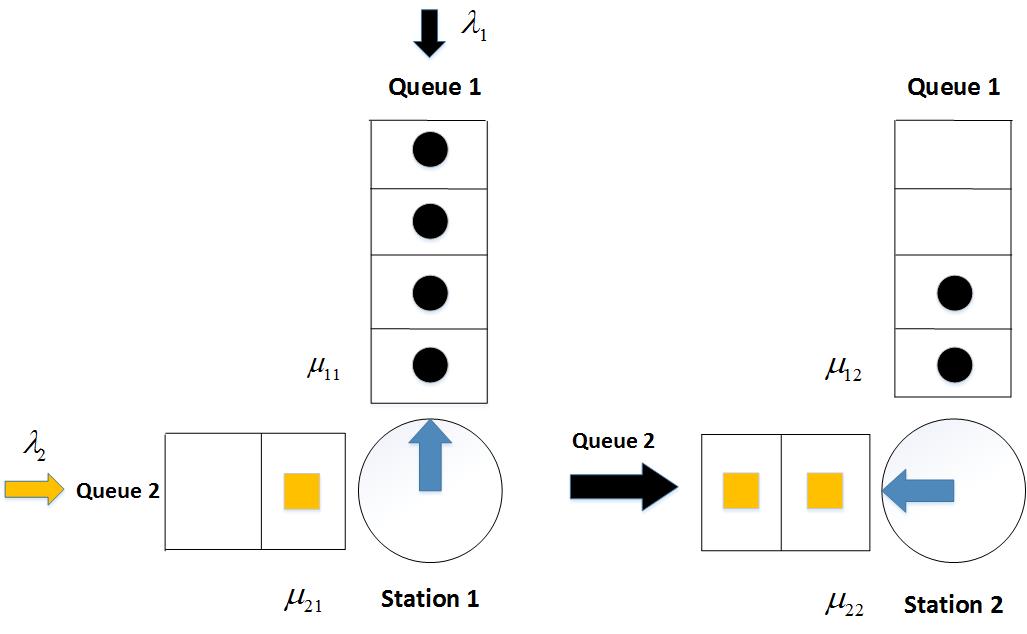}}\label{fig:mesh7.2.1}}%
    \qquad
    \subfloat[Server at station 1 is serving type 2 products while server at station 2 is serving type 1 products.]{{\includegraphics[width=7.5cm, frame]{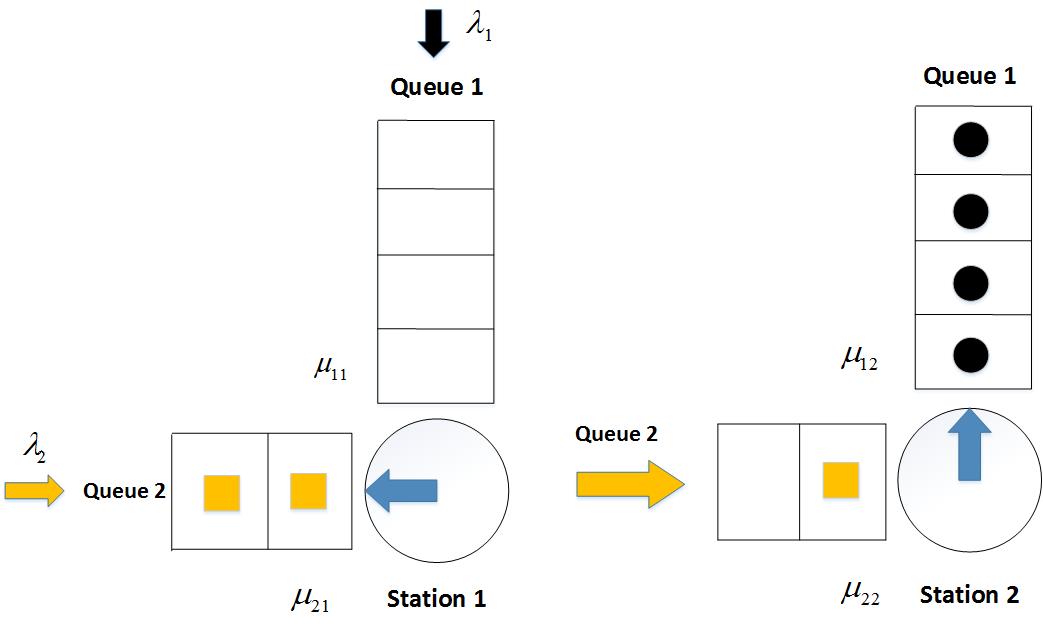}}\label{fig:mesh7.2.2}}%
    \caption{Illustration of out-of-sync polling strategy (OP).}%
    \label{fig:mesh7.2}%
\end{figure}

\noindent \textbf{Out-of-sync polling strategy (OP)}$\colon$ In this strategy, the server at station 1 operates under exhaustive policy, but the server at station 2 always serves a product type different from the product type being served at station 1. In other words, if the server of station 1 switches to do a setup for products of type 2 after emptying the queue of product type 1, the server at station 2 switches to do setups for products of type 1 even if its queue for type 2 product is not empty. However, if station 1 is empty (in both the queues), station 2 continues to serve its queue until its queue is empty or until station 1 begins to perform setup for a queue again after the arrival of a product. In this event, server at station 2 switches to setup for the other product type. Figure $\ref{fig:mesh7.2}$ illustrates a possible scenario under the out-of-sync polling strategy. Initially (in Figure \ref{fig:mesh7.2.1}), the server at station 1 (2) is serving products of type 1 (2). Subsequently, when the server at station 1 empties its queue, it switches to do setup for products of type 2, and the server at station 2 switches to do a setup for products of type 1, even if the queue of product of type 2 is not empty (in Figure \ref{fig:mesh7.2.2}).\\

\color{black}

In tandem networks of polling queues, coordination of service across different stations is essential for achieving efficient network performance in terms of throughput, queue lengths, and waiting times. One extreme, among the spectrum of coordination mechanisms, is the SP strategy, where we have same operations at consecutive stations so that same product type is processed at the two stations. For example, in the manufacturing of rolled aluminum products, aluminum plates need to be stretched soon after they have been heat treated, before they cool down. This necessitates that both stations process a particular type of alloy at the same time in their respective cycles. In this case, these operations can be modeled as tandem polling queues operating under SP strategy. Another possibility is that downstream operations are deliberately kept out-of-sync with the upstream operations to improve the flow. For example, it may be desired that while the saws are processing one type of aluminum alloy, the downstream inspection station is inspecting the plates processed at the saw earlier. In this case, these operations can be modeled as tandem polling queues operating under OP strategy. Furthermore, to reduce cost, manufacturers often process these plates in batches, and use an exhaustive policy, i.e, serve all products waiting in a queue before switching over to another product type. This practice is adopted to save on setup times. Thus, determining the impact of setup times on waiting times is of key interest.\\

The goal of this paperr is to compare the two strategies based on the following system performance measures for a given maximum buffer level $N_{i}\colon$ (i) average system throughput, $\mathbb{E}\left[TH_{i}\right]$, defined as the average number of parts of product type $i$ produced by the system in a time unit $\left[\text{parts/time unit}\right]$, (ii) average queue length, $\mathbb{E}\left[L_{ij}\right]$, defined as the average amount of material stored in buffer for product type $i$ at station $j \left[\text{parts}\right]$, and (iii) average waiting time, $\mathbb{E}\left[W_{i}\right]$, defined as the average time required by products to go through station 1 and 2 $\left[\text{time units}\right]$.\\

To solve the system described above exactly using the conventional Markov chain $\left(\text{MC}\right)$ approach, we would need to use six-tuple state space resulting in over 0.6 million states for a system for a buffer size of 20 for either of the polling strategy. To address this curse-of-dimensionality, we propose a decomposition technique to analyze the system. We provide the overview of the approach below and provide the details in Section \ref{QueueingAnalysis7}.
\subsection{Overview of the Approach}\label{Approach7}
The main idea is to decompose the system into two subsystems$\colon SS\left(i\right)$  for $i ={}1, 2$ and study each subsystem independently. Subsystem $SS\left(i\right)$ comprise of products of type 1 and type 2 at station 1 and only product $i$ at station 2. We use subsystem $SS\left(i\right)$ to obtain exact estimates of the performance measures for type $i$ products at station 1 and station 2. In solving the subsystem $SS\left(i\right)$, we make use of the fact that the server at station 1 drives the corresponding coordination strategy at station 2, i.e., the server at station 2 makes the switch from one product to another at the same time when the server at station 1 makes the switch. Therefore, we keep track of the queue lengths for both the product type at station 1 and the queue length of only product type $i$ at station 2 when we need to compute the performance measures for product $i$. This decomposition helps reduce the state complexity of the system and yet obtain exact solutions for the performance measures of station 1 and station 2.
\section{Queueing Analysis}\label{QueueingAnalysis7}
In this section, we present the queueing analysis for SP strategy in subsection \ref{SPStrategy}. Then, we describe how the queueing analysis can be extended to analyze OP strategy in subsection \ref{OPStrategy}. Finally, we propose solution algorithm to estimate system throughput, expected queue lengths, and expected waiting times.
\subsection{SP Strategy}\label{SPStrategy}
The state of the subsystem $SS\left(i\right)$ with SP strategy at a given time epoch forms a continuous time Markov chain defined by the tuple $\Big(\,l_{11}, l_{21}, r_{i}, l_{12}\,\Big)$, where $l_{ij}$ is the number of products of type $i$ at station $j$, and $r_{i}$ takes value of $S_{i}$ or $U_{i}$, for $i = 1, 2$, depending on if it is doing a setup for product $i$ or is processing product $i$ respectively. Note that $l_{ij}$ is assumed to be positive integer and 0 when stated. Let $q\Big[\left(l_{11}, l_{21}, r_{i}, l_{12}\right), \left(l_{11}', l_{21}', r_{i}', l_{12}\right)\Big]$ denote the transitions from the state $\left(l_{11}, l_{21}, r_{i}, l_{12}\right)$ to the state $\left(l_{11}', l_{21}', r_{i}', l_{12}\right)$  for $\left(r_{i}, r_{i}'\right) \in \{S_{1}, S_{2}, U_{1}, U_{2}\}$. The transitions for the subsystem $SS\left(i\right)$ with SP strategy are summarized below in Table \ref{Table:7.1}.\\
{\renewcommand{\arraystretch}{0.6}
\begin{table}[H]
\centering
\caption{Transitions for subsystem $SS\left(i\right)$ with SP strategy.}\label{Table:7.1}
\begin{tabular}{| C{4cm} |  C{5cm} |  C{2cm} | C{3.8cm}|}
\hline
\textbf{From state} & \textbf{To state} & \textbf{Condition} & \textbf{Transition rate out}\\
\hline
$\left(l_{11}, l_{21}, S_{i}, l_{12}\right)$ & $\left(l_{11}, l_{21}, S_{i'}, l_{12}\right)$ & $l_{i1} ={} 0$ & \multirow{2}{*}{$\mu_{s_{i}}$}\\
$\left(l_{11}, l_{21}, S_{i}, l_{12}\right)$ & $\left(l_{11}, l_{21}, U_{i}, l_{12}\right)$ &  $l_{i1} > 0$ & \\
\hline
$\left(l_{11}, l_{21}, U_{1}, l_{12}\right)$ & $\left(0, l_{21}, S_{2}, l_{12}+1\right)$ & $l_{11} ={} 1$ & \multirow{2}{*}{$\mu_{11}$}\\
$\left(l_{11}, l_{21}, U_{1}, l_{12}\right)$ & $\left(l_{11}-1, l_{21}, U_{1}, l_{12}+1\right)$ &  $l_{11} > 1$ & \\
\hline
$\left(l_{11}, l_{21}, U_{2}, l_{12}\right)$ & $\left(l_{11}, 0, S_{1}, l_{12}\right)$ & $l_{21} ={} 1$ & \multirow{2}{*}{$\mu_{21}$}\\
$\left(l_{11}, l_{21}, U_{2}, l_{12}\right)$ & $\left(l_{11}, l_{21}-1, U_{2}, l_{12}\right)$ &  $l_{21} > 1$ & \\
\hline
$\left(l_{11}, l_{21}, U_{1}, l_{12}\right)$ & $\left(l_{11}, l_{21}, U_{1}, l_{12}-1\right)$ & $l_{12} \geq {} 1$ & $\mu_{12}$\\
\hline
$\left(l_{11}, l_{21}, S_{1}, l_{12}\right)$ & $\left(l_{11}+1, l_{21}, S_{1}, l_{12}\right)$ & \multirow{4}{*}{--} & \multirow{4}{*}{$\lambda_{1}$}\\
$\left(l_{11}, l_{21}, S_{2}, l_{12}\right)$ & $\left(l_{11}+1, l_{21}, S_{2}, l_{12}\right)$ & &\\
$\left(l_{11}, l_{21}, U_{1}, l_{12}\right)$ & $\left(l_{11}+1, l_{21}, U_{1}, l_{12}\right)$ & &\\
$\left(l_{11}, l_{21}, U_{2}, l_{12}\right)$ & $\left(l_{11}+1, l_{21}, U_{2}, l_{12}\right)$ & &\\
\hline
$\left(l_{11}, l_{21}, S_{1}, l_{12}\right)$ & $\left(l_{11}, l_{21}+1, S_{1}, l_{12}\right)$ & \multirow{4}{*}{--} & \multirow{4}{*}{$\lambda_{2}$}\\
$\left(l_{11}, l_{21}, S_{2}, l_{12}\right)$ & $\left(l_{11}, l_{21}+1, S_{2}, l_{12}\right)$ & &\\
$\left(l_{11}, l_{21}, U_{1}, l_{12}\right)$ & $\left(l_{11}, l_{21}+1, U_{1}, l_{12}\right)$ & &\\
$\left(l_{11}, l_{21}, U_{2}, l_{12}\right)$ & $\left(l_{11}, l_{21}+1, U_{2}, l_{12}\right)$ & &\\
\hline
\end{tabular}
\end{table}

Let $\pi\left(l_{11}, l_{21}, r_{i}, l_{12}\right)$ be the steady-state probability of state $\left(l_{11}, l_{21}, r_{i}, l_{12}\right)$. The Chapman-Kolmogorov equations for the Markov chain for subsystem $SS\left(i\right)$ with SP strategy to and from states $\left(l_{11}, l_{21}, S_{1}, l_{12}\right)$ and $\left(l_{11}, l_{21}, U_{1}, l_{12}\right)$ are given by Equations $\left(\ref{eqn7.1}\right) - \left(\ref{eqn7.12}\right)$.
\setlength{\abovedisplayskip}{0pt}
\setlength{\belowdisplayskip}{0pt}
\setlength{\abovedisplayshortskip}{0pt}
\setlength{\belowdisplayshortskip}{0pt}
\begin{align*}
\intertext{For $l_{11} ={}0, l_{21} ={}0, l_{12} \geq 0\colon$}
\left(\lambda_{1}+\lambda_{2}+\mu_{s_{1}}\right)\pi\left(0, 0, S_{1}, 0\right) = {}\mu_{s_{2}}\pi\left(0, 0, S_{2}, l_{12}\right)+\mu_{21}\pi\left(0, 1, U_{2}, l_{12}\right)\numberthis\label{eqn7.1}\\
\intertext{For $l_{11} ={}0, l_{21} >{}0, l_{12} \geq 0\colon$}
\left(\lambda_{1}+\lambda_{2}+\mu_{s_{1}}\right)\pi\left(0, l_{21}, S_{1}, l_{12}\right) = {}\lambda_{2}\pi\left(0, l_{21}-1, S_{1}, l_{12}\right)\numberthis\label{eqn7.2}\\
\intertext{For $l_{11} >{}0, l_{21} ={}0, l_{12}\geq 0\colon$}
\left(\lambda_{1}+\lambda_{2}+\mu_{s_{1}}\right)\pi\left(l_{11}, 0, S_{1}, l_{12}\right) = {}\lambda_{1}\pi\left(l_{11}-1, 0, S_{1}, l_{12}\right)+\mu_{s_{2}}\pi\left(l_{11}, 0, S_{2}, l_{12}\right)+\mu_{21}\pi\left(l_{11}, 1, U_{2}, l_{12}\right)\numberthis\label{eqn7.3}\\
\intertext{For $l_{11} >{}0, l_{21} >{}0, l_{12}\geq 0\colon$}
\left(\lambda_{1}+\lambda_{2}+\mu_{s_{1}}\right)\pi\left(l_{11}, l_{21}, S_{1}, l_{12}\right) = {}\lambda_{1}\pi\left(l_{11}-1, l_{21}, S_{1}, l_{12}\right)+\lambda_{2}\pi\left(l_{11}, l_{21}-1, S_{1}, l_{12}\right)\numberthis\label{eqn7.4}
\end{align*}
\begin{align*}
\intertext{For $l_{11} ={}1, l_{21} ={}0, l_{12}={}0\colon$}
\left(\lambda_{1}+\lambda_{2}+\mu_{11}\right)\pi\left(1, 0, U_{1}, 0\right)= {}\mu_{s_{1}}\pi\left(1, 0, S_{1}, 0\right)+\mu_{12}\pi\left(1, 0, U_{1}, 1\right)\numberthis\label{eqn7.5}\\
\intertext{For $l_{11} ={}1, l_{21} ={}0, l_{12}>{}0\colon$}
\left(\lambda_{1}+\lambda_{2}+\mu_{11}+\mu_{12}\right)\pi\left(1, 0, U_{1}, l_{12}\right)=\\
{}\mu_{s_{1}}\pi\left(1, 0, S_{1}, l_{12}\right)+\mu_{11}\pi\left(2, 0, U_{1}, l_{12}-1\right)+\mu_{12}\pi\left(1, 0, U_{1}, l_{12}+1\right)\numberthis\label{eqn7.6}\\
\intertext{For $l_{11} ={}1, l_{21} >{}0, l_{12}={}0\colon$}
\left(\lambda_{1}+\lambda_{2}+\mu_{11}\right)\pi\left(1, l_{21}, U_{1}, 0\right)= {}\mu_{s_{1}}\pi\left(1, l_{21}, S_{1}, 0\right)+\lambda_{2}\left(1, l_{21}-1, U_{1}, 0\right)+\mu_{12}\pi\left(1, l_{21}, U_{1}, 1\right)\numberthis\label{eqn7.7}\\
\intertext{For $l_{11} ={}1, l_{21} >{}0, l_{12}>{}0\colon$}
\left(\lambda_{1}+\lambda_{2}+\mu_{11}+\mu_{12}\right)\pi\left(1, l_{21}, U_{1}, l_{12}\right)= \\
{}\mu_{s_{1}}\pi\left(1, l_{21}, S_{1}, l_{12}\right)+\lambda_{2}\left(1, l_{21}-1, U_{1}, l_{12}\right)+\mu_{11}\pi\left(2, l_{21}, U_{1}, l_{12}-1\right)+\mu_{12}\pi\left(1, l_{21}, U_{1}, l_{12}+1\right)\numberthis\label{eqn7.8}\\
\end{align*}
\begin{align*}
\intertext{For $l_{11} >{}1, l_{21} ={}0, l_{12}={}0\colon$}
\left(\lambda_{1}+\lambda_{2}+\mu_{11}\right)\pi\left(l_{11}, 0, U_{1}, 0\right)= {}\mu_{s_{1}}\pi\left(l_{11}, 0, S_{1}, 0\right)+\lambda_{1}\left(l_{11}-1, 0, U_{1}, 0\right)+\mu_{12}\pi\left(l_{11}, 0, U_{1}, 1\right)\numberthis\label{eqn7.9}\\
\intertext{For $l_{11} >{}1, l_{21} ={}0, l_{12}>0\colon$}
\left(\lambda_{1}+\lambda_{2}+\mu_{11}+\mu_{12}\right)\pi\left(l_{11}, 0, U_{1}, l_{12}\right)={}\\
\mu_{s_{1}}\pi\left(l_{11}, 0, S_{1}, l_{12}\right)+\lambda_{1}\left(l_{11}-1, 0, U_{1}, l_{12}\right)+\mu_{11}\pi\left(l_{11}+1, 0, U_{1}, l_{12}-1\right)+\mu_{12}\pi\left(l_{11}, 0, U_{1}, l_{12}+1\right)\numberthis\label{eqn7.10}\\
\intertext{For $l_{11} >{}1, l_{21} >{}0, l_{12}={}0\colon$}
\left(\lambda_{1}+\lambda_{2}+\mu_{11}\right)\pi\left(l_{11}, l_{21}, U_{1}, 0\right) = {}\\
\mu_{s_{1}}\pi\left(l_{11}, l_{21}, S_{1}, 0\right)+\lambda_{1}\left(l_{11}-1, l_{21}, U_{1}, 0\right)+\lambda_{2}\left(l_{11}, l_{21}-1, U_{1}, 0\right)+\mu_{12}\pi\left(l_{11}, l_{21}, U_{1}, 1\right)\numberthis\label{eqn7.11}\\
\intertext{For $l_{11} >{}1, l_{21} >{}0, l_{12} > 0\colon$}
\left(\lambda_{1}+\lambda_{2}+\mu_{11}+\mu_{12}\right)\pi\left(l_{11}, l_{21}, U_{1}, l_{12}\right) = {}\\
\mu_{s_{1}}\pi\left(l_{11}, l_{21}, S_{1}, l_{12}\right)+\lambda_{1}\left(l_{11}-1, l_{21}, U_{1}, l_{12}\right)+\lambda_{2}\left(l_{11}, l_{21}-1, U_{1}, l_{12}\right)+\\
\mu_{11}\pi\left(l_{11}+1, l_{21}, U_{1}, l_{12}-1\right)+\mu_{12}\pi\left(l_{11}, l_{21}, U_{1}, l_{12}+1\right)\numberthis\label{eqn7.12}
\end{align*}
Similarly, we can write the balance equations for states of the form $\left(l_{11}, l_{21}, S_{2}, l_{12}\right)$ and $\left(l_{11}, l_{21}, U_{2}, l_{12}\right)$. The normalization condition is written as$\colon$
\begin{align*}
\mathop{\sum\sum\sum\sum}_{\substack{S_{i}\in\{S_{1}, S_{2}\}\\
\left(l_{11}, l_{21}, l_{12}\right)\in\mathbb{Z}}}\pi\left(l_{11}, l_{21}, S_{i}, l_{12}\right)
+\mathop{\sum\sum\sum\sum}_{\substack{U_{i}\in\{U_{1}, U_{2}\}\\
l_{i1}\in\mathbb{Z}^{+}, \text{ }\left(l_{i'1}, l_{i2}\right)\in\mathbb{Z}}}\pi\left(l_{11}, l_{21}, U_{i}, l_{12}\right)= {}1\numberthis\label{eqn7.13}\\
\end{align*}

Using Equations $\left(\ref{eqn7.1}\right)-\left(\ref{eqn7.13}\right)$, we obtain the values of all steady state probabilities for subsystem $SS\left(i\right)$ for $i={}1, 2$ when operating under the SP strategy. We independently analyze subsystem $SS\left(i'\right)$ with SP strategy to obtain the values of all steady state probabilities to estimate performance measures of product $i'$. Using the steady state probabilities, we obtain expressions for average throughput $TH_{ij}$, average queue length $L_{ij}$, and average waiting time $W_{ij}$, of product type $i$, for $i ={}1, 2$ at station j, for $j ={}1, 2$ and are given by Equation $\left(\ref{eqn7.14}\right)$, Equation $\left(\ref{eqn7.15}\right)$, and Equation $\left(\ref{eqn7.16}\right)$ respectively.
\begin{align*}
TH_{i1} &= {}\mu_{i1}\mathop{\sum\sum\sum}\limits_{l_{11}\in\mathbb{Z}^{+}, \text{ }\left(l_{21}, l_{12}\right)\in\mathbb{Z}}\pi\left(l_{11}, l_{21}, U_{i}, l_{12}\right)\\\numberthis\label{eqn7.14}
TH_{i2} &= {}\mu_{i2}\mathop{\sum\sum\sum}\limits_{\left(l_{11}, l_{12}\right)\in\mathbb{Z}^{+}, \text{ } l_{21}\in\mathbb{Z}}\pi\left(l_{11}, l_{21}, U_{i}, l_{12}\right) & i ={} 1, 2.
\end{align*}

\begin{align*}
L_{i1} &= {}\mathop{\sum\sum\sum\sum}_{\substack{r\in\{{S_{1}, S_{2}, U_{1}, U_{2}\}}\\
\left(l_{11}, l_{21}, l_{i2}\right)\in\mathbb{Z}}}
l_{i1}\cdot\pi\left(l_{11}, l_{21}, r, l_{i2}\right)\\\numberthis\label{eqn7.15}
L_{i2} &= {}\mathop{\sum\sum\sum\sum}_{\substack{r\in\{{S_{1}, S_{2}, U_{1}, U_{2}\}}\\
\left(l_{11}, l_{21}, l_{i2}\right)\in\mathbb{Z}}}
l_{i1}\cdot\pi\left(l_{11}, l_{21}, r, l_{i2}\right) & i ={} 1, 2.\\
W_{ij}& = {}L_{ij}TH_{ij}^{-1} & i ={} 1, 2.\numberthis\label{eqn7.16}
\end{align*}

The mean waiting time for product $i$, for $i ={}1, 2$ in the system is given by Equation $\left(\ref{eqn7.17}\right)$.
\begin{align*}
W_{i} &= {}W_{i1}+W_{i2} & i ={} 1, 2. \numberthis\label{eqn7.17}
\end{align*}
\subsection{OP Strategy}\label{OPStrategy}
The state of the subsystem $SS\left(i\right)$ with OP strategy at a given time epoch forms a continuous time Markov chain defined by the tuple $\Big(\,l_{11}, l_{21}, r_{i}, l_{12}\,\Big)$. This state-space representation for OP strategy is identical to state-space representation of SP strategy. The transitions for the subsystem $SS\left(i\right)$ with OP strategy are also same for all except one instance. This difference in the transition is for the service of products at station 2, i.e., the transition $q\Big[\left(l_{11}, l_{21}, U_{2}, l_{12}\right), \left(l_{11}, l_{21}, U_2, l_{12}-1\right)\Big]$. This transition in OP strategy occurs with rate $\mu_{12}$.\\

Let $\pi\left(l_{11}, l_{21}, r_{i}, l_{12}\right)$ be the steady-state probability of state $\left(l_{11}, l_{21}, r_{i}, l_{12}\right)$. The Chapman-Kolmogorov equations for the Markov chain for subsystem $SS\left(i\right)$ unique to OP strategy to and from states $\left(l_{11}, l_{21}, S_{1}, l_{12}\right)$ and $\left(l_{11}, l_{21}, U_{1}, l_{12}\right)$ are given by Equations $\left(\ref{eqn7.18}\right) - \left(\ref{eqn7.25}\right)$. The Equations $\left(\ref{eqn7.1}\right) - \left(\ref{eqn7.4}\right)$ are also valid for SP strategy.
\setlength{\abovedisplayskip}{0pt}
\setlength{\belowdisplayskip}{0pt}
\setlength{\abovedisplayshortskip}{0pt}
\setlength{\belowdisplayshortskip}{0pt}
\begin{align*}
\intertext{For $l_{11} ={}1, l_{21} ={}0, l_{12}={}0\colon$}
\left(\lambda_{1}+\lambda_{2}+\mu_{11}\right)\pi\left(1, 0, U_{1}, 0\right)= {}\mu_{s_{1}}\pi\left(1, 0, S_{1}, 0\right)\numberthis\label{eqn7.18}\\
\intertext{For $l_{11} ={}1, l_{21} ={}0, l_{12}>{}0\colon$}
\left(\lambda_{1}+\lambda_{2}+\mu_{11}\right)\pi\left(1, 0, U_{1}, l_{12}\right)=\\
{}\mu_{s_{1}}\pi\left(1, 0, S_{1}, l_{12}\right)+\mu_{11}\pi\left(2, 0, U_{1}, l_{12}-1\right)\numberthis\label{eqn7.19}\\
\intertext{For $l_{11} ={}1, l_{21} >{}0, l_{12}={}0\colon$}
\left(\lambda_{1}+\lambda_{2}+\mu_{11}\right)\pi\left(1, l_{21}, U_{1}, 0\right)= {}\mu_{s_{1}}\pi\left(1, l_{21}, S_{1}, 0\right)+\lambda_{2}\left(1, l_{21}-1, U_{1}, 0\right)\numberthis\label{eqn7.20}\\
\intertext{For $l_{11} ={}1, l_{21} >{}0, l_{12}>{}0\colon$}
\left(\lambda_{1}+\lambda_{2}+\mu_{11}\right)\pi\left(1, l_{21}, U_{1}, l_{12}\right)= \\
{}\mu_{s_{1}}\pi\left(1, l_{21}, S_{1}, l_{12}\right)+\lambda_{2}\left(1, l_{21}-1, U_{1}, l_{12}\right)+\mu_{11}\pi\left(2, l_{21}, U_{1}, l_{12}-1\right)\numberthis\label{eqn7.21}\\
\intertext{For $l_{11} >{}1, l_{21} ={}0, l_{12}={}0\colon$}
\left(\lambda_{1}+\lambda_{2}+\mu_{11}\right)\pi\left(l_{11}, 0, U_{1}, 0\right)= {}\mu_{s_{1}}\pi\left(l_{11}, 0, S_{1}, 0\right)+\lambda_{1}\left(l_{11}-1, 0, U_{1}, 0\right)\numberthis\label{eqn7.22}\\
\intertext{For $l_{11} >{}1, l_{21} ={}0, l_{12}>0\colon$}
\left(\lambda_{1}+\lambda_{2}+\mu_{11}\right)\pi\left(l_{11}, 0, U_{1}, l_{12}\right)={}\\
\mu_{s_{1}}\pi\left(l_{11}, 0, S_{1}, l_{12}\right)+\lambda_{1}\left(l_{11}-1, 0, U_{1}, l_{12}\right)+\mu_{11}\pi\left(l_{11}+1, 0, U_{1}, l_{12}-1\right)\numberthis\label{eqn7.23}\\
\intertext{For $l_{11} >{}1, l_{21} >{}0, l_{12}={}0\colon$}
\left(\lambda_{1}+\lambda_{2}+\mu_{11}\right)\pi\left(l_{11}, l_{21}, U_{1}, 0\right) = {}\\
\mu_{s_{1}}\pi\left(l_{11}, l_{21}, S_{1}, 0\right)+\lambda_{1}\left(l_{11}-1, l_{21}, U_{1}, 0\right)+\lambda_{2}\left(l_{11}, l_{21}-1, U_{1}, 0\right)\numberthis\label{eqn7.24}\\
\intertext{For $l_{11} >{}1, l_{21} >{}0, l_{12} > 0\colon$}
\left(\lambda_{1}+\lambda_{2}+\mu_{11}\right)\pi\left(l_{11}, l_{21}, U_{1}, l_{12}\right) = {}\\
\mu_{s_{1}}\pi\left(l_{11}, l_{21}, S_{1}, l_{12}\right)+\lambda_{1}\left(l_{11}-1, l_{21}, U_{1}, l_{12}\right)+\lambda_{2}\left(l_{11}, l_{21}-1, U_{1}, l_{12}\right)+\\
\mu_{11}\pi\left(l_{11}+1, l_{21}, U_{1}, l_{12}-1\right)\numberthis\label{eqn7.25}
\end{align*}
The normalization equation in OP strategy is similar to the normalization equation for SP and is given by Equation $\left(\ref{eqn7.13}\right)$. Using Equations $\left(\ref{eqn7.18}\right)-\left(\ref{eqn7.25}\right)$, we obtain the values of all steady state probabilities for subsystem $SS\left(i\right)$ for $i={}1, 2$ when operating under the OP strategy. We independently analyze subsystem $SS\left(i'\right)$ with OP strategy to obtain the values of all steady state probabilities to estimate performance measures of product $i'$. Using the steady state probabilities, we obtain expressions for average throughput $TH_{ij}$, average queue length $L_{ij}$, and average waiting time $W_{ij}$, of product type $i$, for $i ={}1, 2$ at station j, for $j ={}1, 2$ and are given by Equation $\left(\ref{eqn7.14}\right)$, Equation $\left(\ref{eqn7.15}\right)$, and Equation $\left(\ref{eqn7.16}\right)$ respectively.
\section{Comparison of SP and OP Strategy}\label{NumericalResults7}
In this section, we discuss the results of numerical experiments conducted to compare the performance of polling queues under SP and OP strategies. We conduct three sets of experiments pertaining to four different settings. In the first set, we analyze the performance of the two strategies under a symmetric setting. In the second set, we investigate the effect of station asymmetry on system performance. In the third set of experiment, we investigate the impact of different types of product asymmetry on system performance.

\subsection{Comparison for Symmetric Networks}\label{Performance of Symmetric Networks}
In this experiment setting, we study the effects of variation of buffer size on symmetric tandem polling systems and analyze the performance of the two strategies for a symmetric system. Let $\rho_{{j}_{\mathcal{N}_{1}, \mathcal{N}_{2}}}$ be the traffic intensity at station $j$ when the buffer size considered is $N_{1}$ and $N_{2}$ for type 1 and type 2 products, such that $\rho_{j_{\infty, \infty}}$ represents traffic intensity at a station when $\mathcal{N}_{1} = {} \mathcal{N}_{2} = {} \infty$. We set the arrival rate $\lambda_{i}$ to 1 for both the product types. We vary the service rates and the buffer levels, and the results of this analysis are summarized in Table \ref{T:7.2}. Note that, as we analyze symmetric system under this setting, $W_{1} = {} W_{2}$, $TH_{11} = {}TH_{21}$, and $TH_{12} = {}TH_{22}$.\\
\begin{table}[H]
\centering
\caption{Performance analysis of symmetric systems.}\label{T:7.2}
\begin{tabular}{|C{1.50cm}||C{1.50cm}|C{1.50cm}|C{1.50cm}||C{1.50cm}|C{1.50cm}|C{1.50cm}|}
\hline
\multicolumn{7}{|c|}{\textbf{System parameters}$\colon$ $\lambda_{i} = {}1, \mu_{ij} = {} 4.00, \mu_{s_{j}}={}5.00, \rho_{j_{\infty, \infty}} = {} 0.50$}\\
\hline
\multicolumn{1}{|c||}{Input} & \multicolumn{3}{c||}{\textbf{Synchronous (SP)}} & \multicolumn{3}{c|}{\textbf{Out-of-sync (OP)}}\\
\hline
&&&&&&\\[-1em]
	 $N_{i}$ &  $TH_{i1}$ & $TH_{i2}$ & $W_{i}$ & $TH_{i1}$ & $TH_{i2}$ & $W_{i}$\\
	\hline
	3	&	0.94	&	0.70	&	3.85	&	0.94	&	0.54	&	4.02	\\
	6	&	1.00	&	0.85	&	5.31	&	1.00	&	0.71	&	5.58	\\
	9	&	1.00	&	0.90	&	6.84	&	1.00	&	0.80	&	7.08	\\
	12	&	1.00	&	0.92	&	8.35	&	1.00	&	0.84	&	8.57	\\
	15	&	1.00	&	0.94	&	9.86	&	1.00	&	0.87	&	10.07	\\
	\hline
\multicolumn{7}{|c|}{\textbf{System parameters}$\colon$ $\lambda_{i} = {}1, \mu_{ij} = {} 2.50, \mu_{s_{j}}={}5.00, \rho_{j_{\infty, \infty}} = {} 0.80$}\\
\hline
\multicolumn{1}{|c||}{Input} & \multicolumn{3}{c||}{\textbf{Synchronous (SP)}} & \multicolumn{3}{c|}{\textbf{Out-of-sync (OP)}}\\
\hline
&&&&&&\\[-1em]
	 $N_{i}$ &  $TH_{i1}$ & $TH_{i2}$ & $W_{i}$ & $TH_{i1}$ & $TH_{i2}$ & $W_{i}$\\
	\hline
	3	&	0.85	&	0.64	&	4.61	&	0.85	&	0.47	&	4.97	\\
	6	&	0.95	&	0.81	&	6.43	&	0.95	&	0.61	&	7.23	\\
	9	&	0.98	&	0.88	&	8.18	&	0.98	&	0.70	&	9.16	\\
	12	&	0.99	&	0.91	&	9.85	&	0.99	&	0.77	&	10.91	\\
	15	&	1.00	&	0.93	&	11.45	&	1.00	&	0.81	&	12.55	\\
	\hline
\end{tabular}
\end{table}

As expected, we observe that the mean waiting times $W_{i}$ and systems throughput $TH_{ij}$ increases with increase in buffer size for both the polling strategies. Higher buffers result in lower loss probability of arriving products at each station resulting in more throughput. We also note that the SP strategy outperforms OP strategy in terms of systems throughput and waiting times. One explanation for this performance of OP strategy in terms of throughout is its orthogonal service policy at the two stations. As the server is always serving opposite product type at station 2 in case of OP strategy, the queue length becomes equal to the buffer size at station 2 more often in case of OP strategy leading to higher throughput loss, as opposed to SP strategy. In case of SP strategy, since same product type is served at both the stations, the arriving products at station 2 is processed in the same cycle leading to lower throughput loss. The reason for lower waiting times for SP strategy is that most of the products after receiving service at station 1, do not  need to wait for an additional cycle at station 2 as they are processed in the same cycle. However, in case of OP strategy, the products need to wait for at least an additional half of a cycle thereby increasing the waiting times.
\subsection{Comparison for Networks with Station Asymmetry}\label{Performance of Station Asymmetry}
In this experiment setting, we analyze the impact of station asymmetry by examining the effects of upstream bottlenecks and downstream bottlenecks. To analyze the effect of upstream bottleneck, we set the service rate $\mu_{i1}$ at station 1 to 2.50 and the service rates $\mu_{i2}$ at station 2 for both the products types to 4.00. Under this setting of service rates, we vary the buffer size and record the performance measures. Next, to study the effects of downstream bottlenecks, we set the service rate $\mu_{i1}$ at station 1 to 4.00 and the service rates $\mu_{i2}$ at station 2 for both the products types to 2.50. Under this setting of service rates, we vary the buffer size and record the performance measures. The results of this analysis are summarized in Table \ref{T:7.3}. Since we have only station asymmetry, $W_{1j} = {} W_{2j}$ for $j ={}1, 2$ and $W_{1} = {} W_{2}$.\\
\begin{table}[H]
\centering
\caption{Analysis of impact of upstream bottleneck station.}\label{T:7.3}
\begin{tabular}{|C{1.30cm}||C{1.30cm}|C{1.30cm}|C{1.30cm}|C{1.30cm}||C{1.30cm}|C{1.30cm}|C{1.30cm}|C{1.30cm}|}
\hline
\multicolumn{9}{|c|}{\textbf{System parameters}$\colon$ $\lambda_{i} = {}1, \mu_{i1} = {} 2.50,  \mu_{i2} = {} 4.00, \mu_{s_{ij}}=5.00, \rho_{1_{\infty, \infty}} = {} 0.80, \rho_{2_{\infty, \infty}} = {} 0.50$}\\
\hline
\multicolumn{1}{|c||}{Input} & \multicolumn{4}{c||}{\textbf{Synchronous (SP)}} & \multicolumn{4}{c|}{\textbf{Out-of-sync (OP)}}\\
\hline
&&&&&&&&\\[-1em]
	 $N_{i}$ & $TH_{i2}$ & $W_{i1}$ &  $W_{i2}$ & $W_{i}$ & $TH_{i2}$ & $W_{i1}$ &  $W_{i2}$ & $W_{i}$\\
	\hline
3	&	0.76	&	1.34	&	2.15	&	3.49	&	0.53	&	1.34	&	2.63	&	3.97	\\
6	&	0.93	&	1.88	&	2.32	&	4.20	&	0.70	&	1.88	&	3.65	&	5.54	\\
9	&	0.97	&	2.21	&	2.40	&	4.61	&	0.80	&	2.21	&	4.43	&	6.64	\\
12	&	0.99	&	2.42	&	2.43	&	4.84	&	0.87	&	2.42	&	5.03	&	7.44	\\
15	&	1.00	&	2.54	&	2.43	&	4.96	&	0.91	&	2.54	&	5.48	&	8.02	\\

\hline
\multicolumn{9}{|c|}{\textbf{System parameters}$\colon$ $\lambda_{i} = {}1, \mu_{i1} = {} 4.00,  \mu_{i2} = {} 2.50, \mu_{s_{ij}}=5.00,  \rho_{1_{\infty, \infty}} = {} 0.50, \rho_{2_{\infty, \infty}} = {} 0.80$}\\
\hline
\multicolumn{1}{|c||}{Input} & \multicolumn{4}{c||}{\textbf{Synchronous (SP)}} & \multicolumn{4}{c|}{\textbf{Out-of-sync (OP)}}\\
\hline
&&&&&&&&\\[-1em]
	 $N_{i}$ & $TH_{i2}$ & $W_{i1}$ &  $W_{i2}$ & $W_{i}$ & $TH_{i2}$ & $W_{i1}$ &  $W_{i2}$ & $W_{i}$\\
	\hline
3	&	0.52	&	0.80	&	4.84	&	5.64	&	0.52	&	0.80	&	3.97	&	4.77	\\
6	&	0.61	&	0.88	&	8.48	&	9.36	&	0.60	&	0.88	&	7.26	&	8.14	\\
9	&	0.62	&	0.90	&	12.86	&	13.76	&	0.61	&	0.90	&	11.11	&	12.01	\\
12	&	0.62	&	0.90	&	17.51	&	18.41	&	0.62	&	0.90	&	15.30	&	16.20	\\
15	&	0.62	&	0.90	&	22.27	&	22.27	&	0.62	&	0.90	&	19.72	&	20.62	\\
\hline
\end{tabular}
\end{table}

Table \ref{T:7.3} show that SP strategy performs better than OP strategy in terms of waiting times and throughput when upstream station is the bottleneck. However, when downstream station is the bottleneck, OP strategy outperforms SP strategy. We also observe that in the case of bottleneck at downstream station, the throughput converges to a value which is less than the arrival rate to the system. This is because bottleneck at downstream station in synchronization leads to instability in the queue. As a result of the instability, there is no gain in systems throughput with the increase in buffer size.
\subsection{Comparison for Networks with Product Asymmetry}\label{PerformanceForNetworksWith ProductAsymmetry}
In this experiment setting, we study the effects of product asymmetry. To study this, we fix the service rates $\mu_{1j}$ of type 1 products at both the stations to 2.50 and vary the service rates $\mu_{2j}$ of type 2 products at both the stations such such that $\mu_{1j} / \mu_{2j}$ varies between 0.40 to 0.80. Note that for this setting, product 2 has faster service rate at both the stations. We list the results in Table \ref{T:7.4}.\\
\begin{table}[H]
\centering
\caption{Performance analysis of networks with product asymmetry.}\label{T:7.4}
\begin{tabular}{|C{1.40cm}||C{1.40cm}|C{1.40cm}|C{1.40cm}|C{1.40cm}||C{1.40cm}|C{1.40cm}|C{1.40cm}|C{1.40cm}|}
\hline
\multicolumn{9}{|c|}{$\textbf{System parameters}\colon\lambda_{i} = {}1, \mu_{1j} = {} 2.50,  \mu_{2j} = {} 3.13, \mu_{1j}/  \mu_{2j} = {} 0.80, \mu_{s_{ij}}=5.00, \rho_{j_{\infty, \infty}} = {} 0.72$}\\
\hline
\multicolumn{1}{|c||}{Input} & \multicolumn{4}{c||}{\textbf{Synchronous (SP)}} & \multicolumn{4}{c|}{\textbf{Out-of-sync (OP)}}\\
\hline
&&&&&&&&\\[-1em]
	 $N_{i}$ &  $TH_{12}$ & $TH_{22}$ &  $W_{1}$ & $W_{2}$ &  $TH_{12}$ & $TH_{22}$ &  $W_{1}$ & $W_{2}$\\
	\hline
	3	&	0.66	&	0.65	&	4.31	&	4.44	&	0.44	&	0.52	&	5.36	&	4.15	\\
	6	&	0.84	&	0.82	&	5.93	&	6.14	&	0.59	&	0.70	&	7.86	&	5.76	\\
	9	&	0.89	&	0.89	&	7.55	&	7.76	&	0.67	&	0.80	&	10.23	&	7.01	\\
	12	&	0.92	&	0.92	&	9.12	&	9.32	&	0.71	&	0.87	&	12.67	&	6.15	\\
	15	&	0.94	&	0.94	&	10.66	&	10.85	&	0.74	&	0.91	&	15.24	&	8.88	\\
	\hline
\multicolumn{9}{|c|}{$\textbf{System parameters}\colon\lambda_{i} = {}1, \mu_{1j} = {} 2.50,  \mu_{2j} = {} 6.25, \mu_{1j}/  \mu_{2j} = {} 0.80, \mu_{s_{ij}}=5.00, \rho_{j_{\infty, \infty}} = {} 0.56$}\\
\hline
\multicolumn{1}{|c||}{Input} & \multicolumn{4}{c||}{\textbf{Synchronous (SP)}} & \multicolumn{4}{c|}{\textbf{Out-of-sync (OP)}}\\
\hline
&&&&&&&&\\[-1em]
	 $N_{i}$ &  $TH_{12}$ & $TH_{22}$ &  $W_{1}$ & $W_{2}$ &  $TH_{12}$ & $TH_{22}$ &  $W_{1}$ & $W_{2}$\\
	\hline
3	&	0.69	&	0.68	&	3.94	&	4.16	&	0.31	&	0.65	&	8.47	&	2.83	\\
6	&	0.85	&	0.83	&	5.44	&	5.72	&	0.38	&	0.85	&	14.31	&	3.50	\\
9	&	0.90	&	0.89	&	7.00	&	7.22	&	0.39	&	0.94	&	20.98	&	3.85	\\
12	&	0.92	&	0.92	&	8.53	&	8.72	&	0.40	&	0.98	&	28.14	&	4.03	\\
15	&	0.94	&	0.94	&	10.05	&	10.22	&	0.40	&	0.99	&	35.52	&	4.12	\\
\hline
\end{tabular}
\end{table}
\normalsize
As expected, we observe that the systems throughput $TH_{i2}$ increases with increase in buffer size. Table \ref{T:7.4} shows that for both the coordination strategies, $W_{2}$ (for the product type having the faster service rate) is higher as compared to $W_{1}$. A possible explanation for this is that since the servers at both the stations are faster in serving products of type 2, when they switch to serve products of type 1, because of lower service rates for type 1 products, the server processes products from that queue for a longer duration. As a consequence, the products of type 2 wait longer. We note that the total WIP $\left(\sum\sum_{ij}L_{ij}\right)$ is lower in OP strategy as compared to SP strategy for a given buffer level. The reason for lower WIP in case of OP strategy is higher loss in throughput at station 2 as compared to SP strategy. We also note that $W_{1}$ is lower for SP strategy while $W_{2}$ is lower for OP strategy.
\section{Conclusions}\label{Conclusions7}
In this paper, we analyze the performance of tandem polling queues with finite buffers operating under two different coordination strategies, namely the synchronous polling and and out-of-sync polling strategies. Under Markovian settings, we conduct an exact analysis to determine steady state probabilities, throughput, and mean waiting times for each type of product at each station using a decomposition approach. Recognizing that these strategies could yield substantially different performances, we conduct numerical experiments to evaluate performance under a variety of settings. We study the effect of symmetry and asymmetry in processing characteristics of the products at an individual station and across the stations in the network. We also evaluate performance when stations are balanced and unbalanced in terms of traffic intensities.\\

In all the experiment settings, we observed that the mean waiting times and system throughput increases with the increase in buffer size. In settings with product and station symmetry, we observed that the SP strategy outperforms the OP strategy. In settings with station asymmetry, we observed that SP strategy performs better than OP strategy when the bottleneck is upstream. In case when the bottleneck is downstream, we observe that OP performs better than SP strategy. However, downstream at bottleneck leads to instability in the system. We also observed that in case of product asymmetry, the total WIP is lower in OP strategy as compared to SP strategy for a given buffer level.
\bibliographystyle{nonumber}

\end{document}